\documentclass[12pt]{iopart}
\usepackage[dvips]{graphicx}
\usepackage{amssymb}
\usepackage{bm}
\usepackage{color}
\usepackage[latin1]{inputenc}%permette di usare la tastiera latina

\begin{document}
%\draft

%\hyphenation{a-long}

\title[A view from inside iron-based superconductors]{A view from inside iron-based superconductors}

\author{P Carretta,$^{1}$ R De
Renzi,$^{2}$ G Prando,$^{3}$ S Sanna$^{1}$ }

\address{$^{1}$ Dipartimento di Fisica, Universit\'a di Pavia and
CNISM, I-$27100$ Pavia, Italy}
\address{$^{2}$ Dipartimento di Fisica, Universit\'a di Parma and
CNISM, I-$43124$ Parma, Italy}
\address{$^{3}$ Leibniz-Institut f\"ur Festk\"orper- und
Werkstoffforschung (IFW) Dresden, D-$01171$ Dresden, Germany}

\ead{pietro.carretta@unipv.it}%

%%%%%%%%%%%%%%%%%%%%%%%%%%%%%%%%%%%%%%%%%%%%%%%%%%%%%%%%%%%%%%%%%%%%%%%%%%%%%%%%
\begin{abstract}
Muon spin spectroscopy is one of the most powerful tools to
investigate the microscopic properties of superconductors. In this
manuscript, an overview on some of the main achievements obtained
by this technique in the iron-based superconductors (IBS) are
presented. It is shown how the muons allow to probe the whole
phase diagram of IBS, from the magnetic to the superconducting
phase, and their sensitivity to unravel the modifications of the
magnetic and the superconducting order parameters, as the phase
diagram is spanned either by charge doping, by an external
pressure or by introducing magnetic and non-magnetic impurities.
Moreover, it is highlighted that the muons are unique probes for
the study of the nanoscopic coexistence between magnetism and
superconductivity taking place at the crossover between the two
ground-states.
\end{abstract}
%%%%%%%%%%%%%%%%%%%%%%%%%%%%%%%%%%%%%%%%%%%%%%%%%%%%%%%%%%%%%%%%%%%%%%%%%%%

\pacs{$74$.$70$.Xa, $76$.$75$.+i, $74$.$62$.-c, $74$.$25$.Uv}

\submitto{\PS}

\newpage

\section{Introduction}

\begin{figure}[b!]
\vspace{7cm} \includegraphics{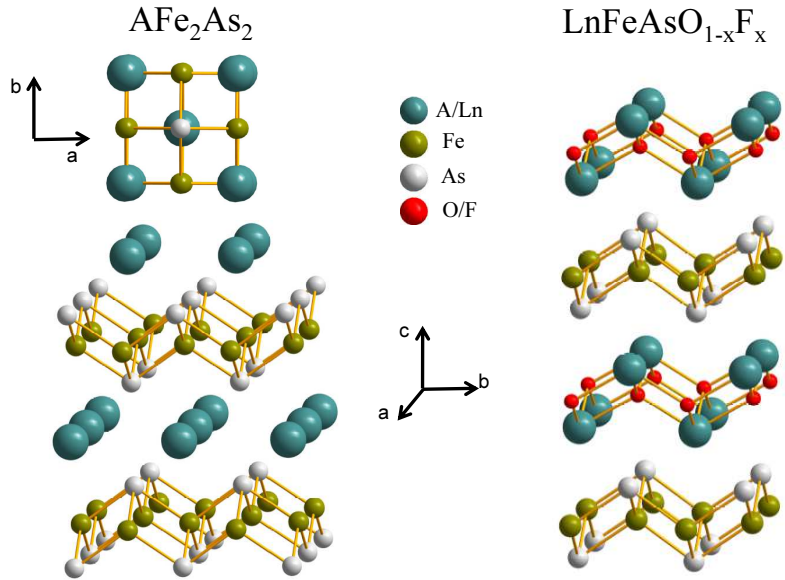} \caption{Crystalline
structures of AFe$_2$As$_2$ (with A alkali ions) (left) and of LnFeAsO$_{1-x}$F$_x$ (with Ln a lanthanide ion)  (right) Fe-based superconductors.
For the AFe$_2$As$_2$ superconductors the unit cell projected along the $c$ axis is also shown at
the top left part of the figure. }\label{Struct_pnict}
\end{figure}

\begin{figure}[h!]
\vspace{8.8cm} \includegraphics{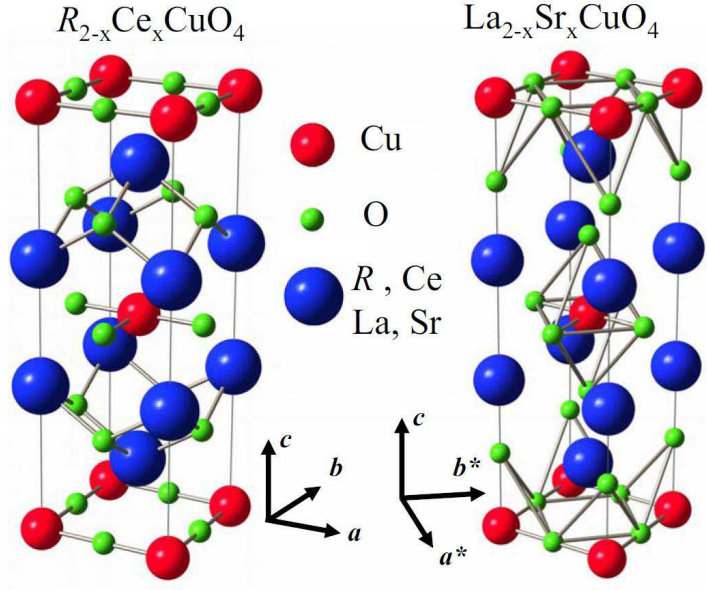} \caption{Crystalline structures
of typical electron- (left) and hole-doped (right) cuprate
materials (Figure reprinted from Ref.\cite{Arm10}. Copyright (2010) by the American Physical Society).}\label{Struct_cupr}
\end{figure}

\begin{figure}[h!]
\vspace{7.7cm} \includegraphics{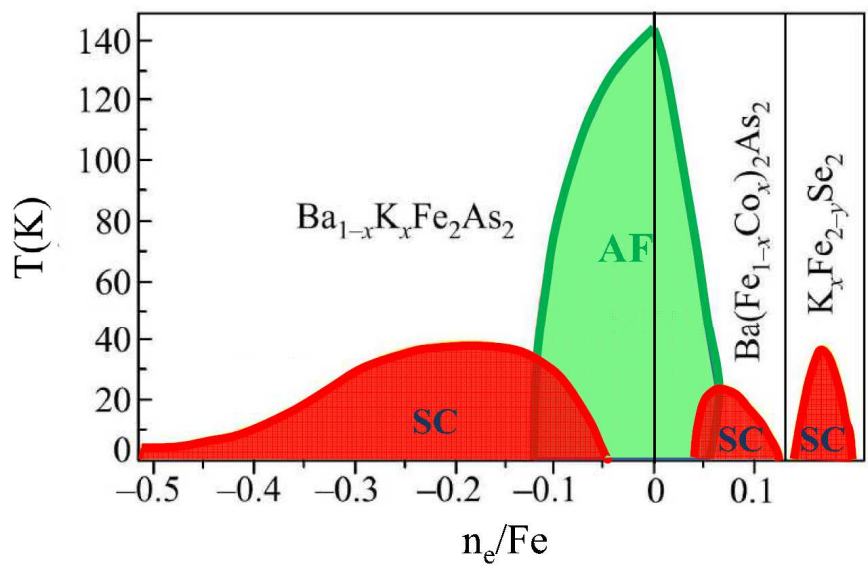} \caption{Typical electronic phase diagram
for the $122$ family of pnictide materials as a function of the number of extra-electrons n$_e$ per Fe atom,
both in the case of hole-
and electron-doping.}\label{PhDiag_pnict}
\end{figure}

\begin{figure}[h!]
\vspace{7.8cm} \includegraphics{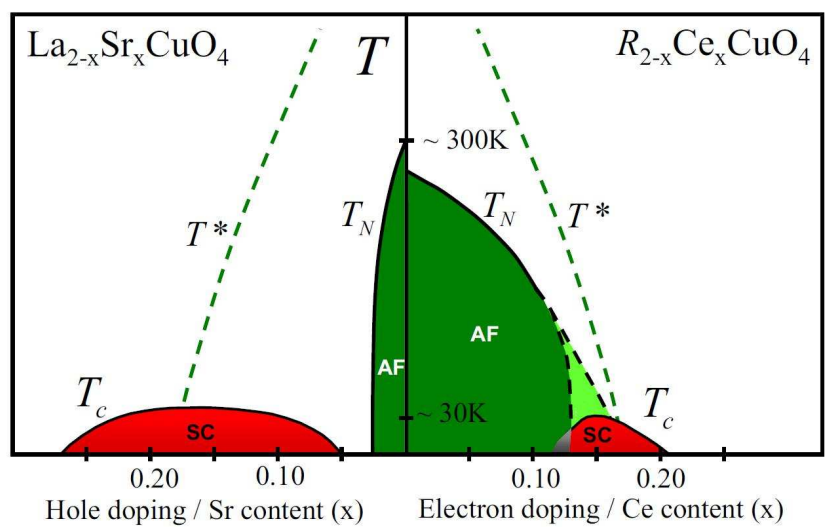} \caption{Typical electronic phase diagram
for cuprate materials both in the case of hole- and electron-doping
(Figure reprinted from Ref. \cite{Arm10}. Copyright (2010) by the American Physical Society).}\label{PhDiag_cupr}
\end{figure}

More than $20$ years after the discovery of high-temperature
superconductivity (HTSC) in the cuprates \cite{Mul86,Lee06}, the
observation of critical temperatures ($T_{c}$) approaching $55$ K
in iron-based pnictides \cite{Kam08,Ren08} has renewed the
interest for superconductivity and for the fascinating
phenomenology it gives rise to \cite{BobWeb}. Remarkably, the
iron-based superconductors (IBS hereafter) share many similarities
with the cuprates \cite{Joh10,Lum10,Pag10,Ste11,Bas11}. First of
all, both families of materials are characterized by layered
structures, involving FePn layers (with Pn = As, P or Se) in the
IBS \cite{Can10} and CuO$_{2}$ planes in the cuprates \cite{Rig98}
(see Figs.~\ref{Struct_pnict} and \ref{Struct_cupr} for the structure of these two
classes of materials, respectively). The layered structure gives
rise to a sizeable anisotropy in the transport and magnetic
properties, which is more marked in the latter compounds. The high
anisotropy is known to cause an enhancement of the flux lines
lattice (FLL) mobility and to yield detrimental dissipative
phenomena \cite{Bla94,Bra95} which could hinder the technological
applicability of these materials. Still, the IBS are characterized by an anisotropy which is not
so pronounced and, accordingly, by a lower FLL
mobility.\cite{Pra11a,Pra12a,Pra12b,Pra13d,Bos13} At the same
time they show upper critical fields often exceeding $60$ Tesla
\cite{Hun08,Fuc09}, making the IBS rather promising for
applications.
The high upper critical fields imply rather short coherence
lengths, reaching few nanometers along the FeAs planes, so that
these materials are weakly sensitive to disorder and, in
particular, to non-magnetic impurities. Another relevant
similarity between the IBS and the cuprates is that
superconductivity is obtained by charge-doping a strongly
correlated electron system with a magnetic ground-state in both
families of compounds \cite{Lee06,Bas11,Maz10}. This is shown in
the typical electronic phase diagrams reported in
Figs.~\ref{PhDiag_pnict} and \ref{PhDiag_cupr} for the IBS and for
the cuprates, respectively.

The correlations are certainly more
relevant in the cuprates where the significant Coulomb repulsion
yields to an insulating ground-state, while in the IBS the
magnetic precursors are metallic. In particular, a multi-band
scenario applies to pnictides where the $5$ bands involving all the Fe $3d$
orbitals contribute to the density of states at the Fermi level
\cite{Sin08a,Sin08b,Sub08}. Notice that a metallic behaviour with
reconstruction of the Fermi surface (FS) is observed also in the
cuprates after a few percent of charge doping, prior to the onset
of the superconducting ground-state \cite{Bat94}.
%and in in the underdoped regime after suppressing the superconducting phase by
%means of a magnetic field \cite{Doi07}.
At the same time, in the pnictides  the scenario of an
antiferromagnetic (AFM) phase arising just from the FS nesting
remains controversial, since non-negligible correlations, bringing
those samples close to a Mott-like transition, should also be
considered \cite{SiQ08,SiQ09,Qaz09,Joh09,Maz09}. As a result,
several authors have described the magnetism in pnictides in terms
of effective Hamiltonians for localized magnetic moments in the
presence of competing frustrating interactions
\cite{SiQ08,SiQ09,Yil08,Sme10,Bon12}.

\begin{figure}[t!]
\vspace{13.4cm} \includegraphics{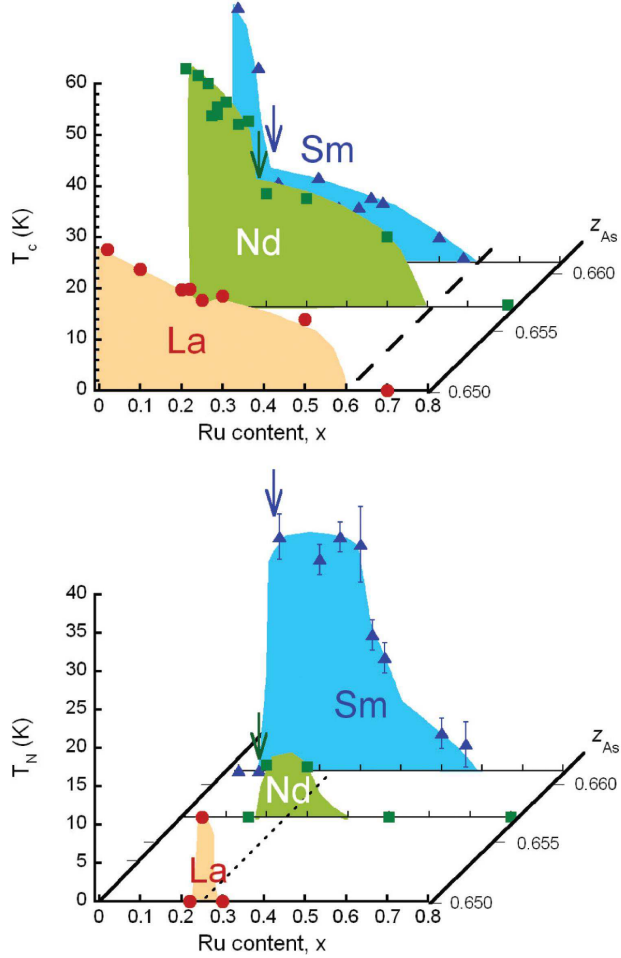} \caption{Electronic phase diagram
for optimally Fluorine-doped LnFeAsO$_{1-y}$F$_y$  $1111$
materials as a function of the amount of Fe$_{1-x}$Ru$_{x}$
dilution, with Ln= Sm, Nd, La \cite{San13}.}\label{Riassuntivo_ru}
\end{figure}

The phase diagrams described above are obtained by doping the
parent compound with charges (either holes or electrons).
Several chemical substitutions have been investigated in order to
achieve such goal. However, as far as the substitution of the Fe
ions with other transition metal (TM) elements is considered, the
effectiveness of charge doping is still under strong debate and
has been subject of recent intense efforts both on the
computational and on the experimental sides, particularly in the
case of Co for Fe substitution. Contrasting results have been
reported from density-functional theory (DFT) calculations
\cite{Wad10,Ber12a} as well as from measurements via x-rays
absorption \cite{Bit11,Mer12} and photoemission \cite{Lev12}
spectroscopies. At the same time, the role of isovalent
substitutions such as Fe$_{1-x}$Ru$_{x}$ attracted great interest
both in LnFeAsO$_{1-x}$F$_x$ (the so-called $1111$ family of IBS,
with Ln a lanthanide ion) and in the AFe$_2$As$_2$ (the $122$ IBS,
with A an alkali ion) compounds. The Fe$_{1-x}$Ru$_{x}$ dilution
does not induce a superconducting phase in LaFeAsO \cite{Bon12},
at variance with what is observed in BaFe$_{2}$As$_{2}$
\cite{Tha10,Dha11}. Still, it induces a reentrant static magnetic
phase nanoscopically coexisting with superconductivity in the
optimally Fluorine-doped $1111$ IBS for different Ln ions
\cite{San11,San13} (see Fig.~\ref{Riassuntivo_ru}).

It should be mentioned that, as an alternative to chemical doping
of the parental compounds, the application of an external
\cite{Tak08,Chu09,Kha11,Gat12} or of a chemical \cite{Wan09}
pressure has been reported to be a useful tool to span through the
different electronic ground states of these materials. However,
the effects of chemical doping and pressures are not equivalent at
all. In particular, the increase of pressure is known to involve
also the modification of more subtle many-body effects like the
Kondo screening in the presence of conduction carriers hybridized
with unpaired $f$ electrons \cite{Pra10a}. It has been established
theoretically that the Kondo effect and superconductivity strongly
compete in $1111$ pnictide compounds \cite{Pou08}. For instance,
this is clearly the case for Ce-based $1111$ compounds where
superconductivity is induced by charge doping
\cite{Shi11,Mae12,Pra13c} but bulk superconductivity cannot be
induced neither by external \cite{Zoc11} nor by chemical
\cite{Jes12} pressures, which are naively expected to enhance the
Kondo coupling.

The similarities among IBS and cuprates discussed above suggested
that the understanding of the microscopic mechanisms at work in
IBS would have eventually lead to the clarification of the long
debated origin of HTSC. The interplay between magnetism and
superconductivity have lead to the idea that in both classes of
materials a pairing  involving the spin excitations could be
present \cite{Pag10,Jin11}. However, in spite of the huge efforts
both on the experimental and theoretical sides, currently it is
not yet clear whether such a pairing mechanism is indeed driving
HTSC. On the other hand, it is well established that for the
hole-doped cuprates the symmetry of the order parameter is d-wave
\cite{Man02}, while it is s-wave for the electron-doped cuprates
\cite{Par08}. In the IBS, the symmetry of the order parameter is
presently subject of an intense debate. While it was initially
suggested that the FS nesting would favour a magnetic coupling
between hole-like and electron-like bands, yielding to an
$s_{\pm}$ symmetry of the order parameter \cite{Maz08,Umm11}, more
recent studies suggested that a conventional $s$ wave symmetry
would be more likely, with a pairing mediated by orbital current
fluctuations \cite{Kon10,Ona12}. Even if the nature of the pairing
mechanism is still undisclosed, it seems to be well established
that a conventional phonon-mediated superconductivity is unlikely
in IBS \cite{Boe08}. Overall, the IBS appear as rather complex
systems, and more than $5$ years after their discovery their
understanding still demands for a significant experimental
investigation and a suitable theoretical modelling.

Muon spin spectroscopy ($\mu^{+}$SR) has played a major role in
the clarification of the electronic and magnetic properties of
cuprate materials first \cite{Uem89,Son07} and of the IBS more
recently. The muons act as nanoscopic Hall sensors which allow to
map the local fields inside those materials and to track their
time evolution. In fact, they allow to probe the spin dynamics,
the local field arising from the onset of a magnetic order or the
field distribution generated by the flux lines in the mixed state
\cite{Son07,Cox87,Dal97,Blu99,Son00,Yao11}. The local nature of
the technique is particularly useful to evidence the intrinsic
microscopic electronic inhomogeneities leading to the nanoscopic
coexistence of magnetic and superconducting domains and/or of
charge poor and charge rich regions, which characterize both the
IBS \cite{San11,Lan10} and the cuprates \cite{Wu11,Gra12}. In this
respect, $\mu^+$SR is rather a unique tool, complementary to
non-local techniques as neutron scattering or to other local
techniques as NMR, which can suffer from RF penetration and/or
excessive line broadening problems. In the following, we present
some of the main achievements obtained by means of $\mu^{+}$SR in
the IBS. First, it is shown how the muons allow to investigate the
whole phase diagram of these materials, from the magnetic to the
superconducting phase, as well as the nature of the phase
transition between the superconducting and magnetic ground-states.
It is then shown how the study of the transverse field relaxation
in the superconducting state provides useful information on the
evolution of the superconducting transition temperature with the
concentration of Cooper pairs. We shall mostly concentrate on the
static properties of IBS by referring to the 1111 family and on
the effect of heterovalent and isovalent chemical substitutions on
its phase diagram and on the superconducting properties. The
scenario observed for the 122 family of IBS will also be briefly
recalled, while for the $\mu$SR studies carried out in other families of IBS, as the 11 chalcogenides
and the 111 the reader can refer to Refs. \cite{R11a,R11b,R11c,R11d,R11e,R11f,R11g} and Ref. \cite{R111}, respectively.

\section{The zero field muon spin polarization in the magnetic phase}\label{SectionZFMuSR}

The full spin-polarization of the positive muon ($\mu^{+}$) beams
produced at meson factories, such as ISIS (Rutherford-Appleton
Laboratories, UK) and S$\mu$S (Paul Scherrer Institut, PSI,
Switzerland), is the prerequisite for running $\mu^{+}$SR
experiments. The possibility to work with fully polarized beams
has the great advantage, in comparison to other local-probe
techniques such as NMR, that there is no need to perturb the
system under investigation with an external polarizing magnetic
field. Accordingly, local magnetism can be investigated even in
conditions of zero-magnetic field (ZF).

In general, at a given temperature ($T$), the ZF $\mu^+$
depolarization as a function of the time ($t$) elapsed after the
implantation into the sample is described by
\begin{equation}\label{EqGeneralFittingZFWithBKG}
    \frac{A_{ZF}(t)}{A_{0}} = \left(1 - a_{s}\right)
    G_{KT}^{H}(t) +
    a_{s} G_{ZF}^{s}(t)\,\,\, .
\end{equation}
Here $A_{0}$ is an instrument-dependent constant corresponding to
the condition of full spin-polarization extrapolated at $t
\rightarrow 0$, $a_{s}$ represents the fraction of $\mu^{+}$
implanted into the investigated sample while $\left(1 -
a_{s}\right)$ is the background fraction. This latter quantity
includes, for example, the $\mu^{+}$ implanted into the sample
holder or into the cryostat walls and probing weakly-magnetic
regions whose main contribution comes from magnetism of nuclear
origin. This typically results in a Gaussian Kubo-Toyabe depolarization
function $G_{KT}^{H}(t)$, which at early times shows a Gaussian trend
$G_{KT}^{H}(t)\sim e^{-\frac{\left(\sigma_{N}\right)^{2}t^{2}}{2}}$,
governed by a weakly $T$-dependent factor $\sigma_{N}$.
Both MuSR and EMU, at ISIS, and GPS and Dolly, at PSI, are
designed as low-background spectrometers, namely allowing one to
achieve the condition $a_{s} \simeq 1$. However, the background
term in Eq.~\ref{EqGeneralFittingZFWithBKG} is of crucial
importance while analyzing data from measurements under applied
pressure performed, for instance, at the GPD facility of S$\mu$S
\cite{Kha11,DeR12,Pra13b}. Here, the thick walls of the pressure
cell are able to stop a sizeable fraction of $\mu^{+}$ leading
typically to $a_{s} \simeq \left(1 - a_{s}\right) \simeq 0.5$.
Nevertheless, since all the measurements to be discussed
subsequently have been performed in the low-background
spectrometers, from now on it is assumed that $a_{s} = 1$, so that
$A_{ZF}(t)/A_0= G^s_{ZF}(t)$, with $G_{ZF}^{s}(t)$ the
spin-depolarization function of the $\mu^{+}$ implanted into the
sample. As it will be explained in the following, in the IBS the
ZF depolarization can be described rather well by
\begin{eqnarray}\label{EqGeneralFittingZF}
    \frac{A_{ZF}(t)}{A_{0}} = G_{ZF}^{s}(t) & = & \left[1 - V_{m}(T)\right]
    G_{KT}^{s}(t)
    {} \nonumber\\ & + & \sum_{i = 1}^{N}
    \zeta_{i} \left[a_{i}^{Tr}(T) F_{i}(t)
    D_{i}^{Tr}(t) + a_{i}^{L}(T)
    D_{i}^{L}(t)\right]\,\,\, .
\end{eqnarray}
%Accordingly, one has
%\begin{eqnarray}\label{EqGeneralFittingZF}
%   \frac{A_{T}(t)}{A_{0}} & = &
%    \left[1 - V_{m}(T)\right]
%    e^{-\frac{\left(\sigma_{N}^{s}\right)^{2}
%    t^{2}}{2}} {} \nonumber\\ & & +
%    \sum_{i = 1}^{N}
%    \zeta_{i} \left[a_{i}^{Tr}(T) F_{i}(t)
%    D_{i}^{Tr}(t) + a_{i}^{L}(T)
%    D_{i}^{L}(t)\right].
%\end{eqnarray}
where $G_{KT}^{s}(t)\simeq e^{-\frac{\left(\sigma_{N}^{s}\right)^2t^2}{2}}$ for $\sigma_{N}^{s}t\ll 1$,
is the Kubo-Toyabe function describing the relaxation arising from the dipolar coupling with the nuclei in the sample.
Eq.~\ref{EqGeneralFittingZF} typically holds in the case of
materials undergoing a phase transition to a magnetically-ordered
state at a temperature $T_{N}$. Accordingly, a set of
ZF-$\mu^{+}$SR measurements at different $T$ values allows one to
access several microscopic quantities associated both with the
ordered and with the paramagnetic phases. The parameter $V_{m}(T)$
represents the fraction of $\mu^{+}$ implanted into the sample and
feeling a spontaneous static magnetic field, namely the sample
magnetic volume fraction, from which one can estimate $T_{N}$. The
ideal case of a step-like behaviour of $V_{m}(T)$ at $T_{N}$ is
modified in real systems where a spatial distribution of
$T_{N}(\bm{r})$ can be present. The assumption of a Gaussian-like
distribution of local transition temperatures generally leads to
the following phenomenological expression for $V_{m}(T)$
\begin{equation}\label{EqMagneVolERFC}
    V_{m}(T) = \frac{1}{2} \; \textrm{erfc}\left[\frac{T -
    T_{N}} {\sqrt{2}\Delta}\right]
\end{equation}
where $\textrm{erfc}(x)$ is the complementary error function.
%is defined as
%\begin{equation}
%    \textrm{erfc}(x) = \frac{2}{\sqrt{\pi}} \int_{x}^{+\infty}
%    e^{-t^{2}} dt.
%\end{equation}
%Eq.~\ref{EqMagneVolERFC} is trivially obtained as the
%convolution of an ideal step-like function (expected in the presence
%of a well-defined $T_{N}$ value) with a Gaussian function allowing
%for a spatial distribution $T_{N}(\textbf{r})$ over the sample
%volume.
A fitting procedure to the experimental $V_{m}(T)$ data
according to Eq.~\ref{EqMagneVolERFC} allows a precise
definition both of $T_{N}$ and of the relative width $\Delta$ of the
transition itself. In particular, the $T_{N}$ value extracted from
Eq.~\ref{EqMagneVolERFC} is an average value defined as the $T$
value corresponding to $50$ \% of the magnetic volume fraction.

Let us now consider the behaviour of the depolarization function
at $T \ll T_{N}$. In general, one should account for the
possibility of different  phases in the sample volume, i.~e. for a
macroscopic or nanoscopic (see Sect. 3) segregation of magnetic
and paramagnetic phases leading to the condition $V_{m}(T) < 1$.
Here, it is assumed $V_{m}(T) = 1$ in the simplified assumption of
a homogeneous fully-magnetic sample. Then the first term in
Eq.~\ref{EqGeneralFittingZF} drops out, leading to
\begin{equation}\label{EqGeneralFittingZFMagnetic}
    \frac{A_{ZF}(t)}{A_{0}} = \sum_{i = 1}^{N} \zeta_{i}
    \left[a_{i}^{Tr}(T)
    F_{i}(t) D_{i}^{Tr}(t) +
    a_{i}^{L}(T) D_{i}^{L}(t)\right].
\end{equation}
The presence of $N$ crystallographically-inequivalent stopping
sites for $\mu^{+}$ is accounted for by the sum over the index
$i$. Each stopping site is characterized by its corresponding
stopping probability $\zeta_{i}$, with $\sum_{i} \zeta_{i} = 1$.
%Even in the presence of a long-range ordered magnetic phase in the
%sample, the implantation in different grains (in the case of
%powders) and even the presence of randomly-oriented magnetic
%domains (in the case of unpolarized magnetic materials) should be
%accounted for.
Accordingly, the quantities $\zeta_{i} a_{i}^{Tr}(T)$ and
$\zeta_{i} a_{i}^{L}(T)$ represent the fractions of all the
implanted $\mu^{+}$ at the site $i$ that feel a static field in
transversal ($Tr$) and longitudinal ($L$) directions with respect
to the initial $\mu^{+}$ polarization, respectively. Thus, the
following relation holds under general assumptions
\begin{equation}
    \sum_{i = 1}^{N} \zeta_{i} \left[a_{i}^{Tr}(T)
    + a_{i}^{L}(T)\right] = V_{m}(T).
\end{equation}
For powder samples, in the ideal condition that long-range
magnetic order develops throughout  the whole sample volume, the
relations
\begin{equation}\label{EqIdealSumOneThirdTwoThird}
    \sum_{i = 1}^{N} \zeta_{i} a_{i}^{Tr}(T \ll
    T_{N}) = \frac{2}{3} \qquad \textrm{and} \qquad
    \sum_{i = 1}^{N} \zeta_{i} a_{i}^{L}(T \ll
    T_{N}) = \frac{1}{3}
\end{equation}
follow from straightforward geometrical arguments. This is
commonly addressed to as the $2/3$ - $1/3$ rule. The function
$F_{i}(t)$ in Eq.~\ref{EqGeneralFittingZFMagnetic} describes the
coherent oscillations associated with the Larmor precession of the
$Tr$ fraction of muon spins around the local magnetic field
$B_{\mu}$, generated by the long-range magnetic order. The
distribution in the magnitude of the local fields give rise to the
corresponding damping functions $D_{i}^{Tr}(t)$. On the other
hand, $\mu^{+}$ belonging to the $L$ fraction do not precess and
usually probe longitudinal $T_{1}$-like dynamical processes
resulting in a relaxing tail described by $D_{i}^{L}(t)$. Due to
the limited time window, to the finite number of counts and to the
comparable values of the relaxation constants, it is often
difficult to distinguish the longitudinal relaxations of different
$\mu^{+}$ sites, which eventually merge in a common $D^{L}(t)$.
Typically, $D^{L}(t)$ is an exponentially-relaxing component
characterized by weak relaxation rates, typically of the order of $0.1 \;
\mu$s$^{-1}$.

In the opposite temperature limit ($T \gg T_{N}$), namely in the
paramagnetic phase, no static fields of electronic origin alter
the initial $\mu^{+}$ polarization. In this case, $V_{m}(T) =
\sum_{i} \zeta_{i} a_{i}^{Tr} = \sum_{i} \zeta_{i} a_{i}^{L} = 0$
and the whole second term in Eq.~\ref{EqGeneralFittingZF}
vanishes, leading to
\begin{equation}\label{EqGeneralFittingZFParamagnetic}
    A_{ZF}(t) \simeq  A_{0} e^{-\frac{\left(\sigma_{N}^{s}\right)^{2}
    t^{2}}{2}}\,\,\, ,
\end{equation}
namely the decay is determined by the dipolar field distribution arising from the nuclear
magnetic moments.

\begin{figure}[t!]
\vspace{9.2cm} \includegraphics{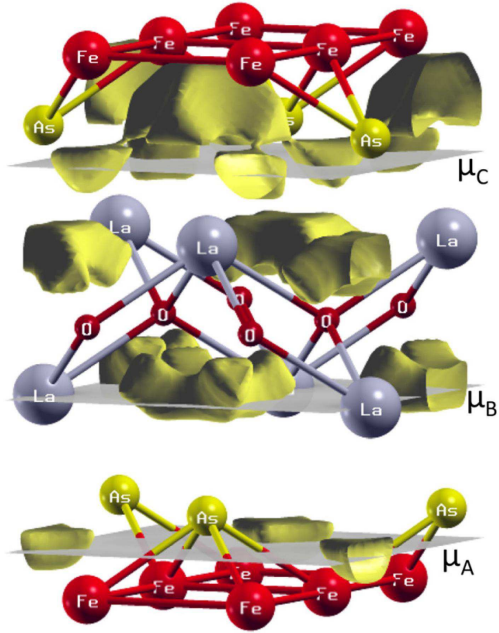} \caption{\label{MuonLandscape}
Electrostatic landscape for LaFeAsO after DFT calculations
 \cite{DeR12}. Iso-potential curves corresponding
to the thermalization positions for $\mu^{+}$ are represented as
yellow shaded regions. Only sites ``A'' and ``B'' are stable minima
while site ``C'' is unstable against the zero-point energy of
$\mu^{+}$.}
\end{figure}
In the case of $1111$ Fe-based oxy-pnictide materials, the
crystallographic sites for the thermalization of $\mu^{+}$ have
been computed by means of DFT calculations of the electrostatic
potential as reported, e.~g., in \cite{DeR12,Mae09} (see also
\cite{Pra13b} for the case of RECoPO samples). The results of the
calculations yield $N = 2$ stable minima in addition to a further
site which, however, is unstable against the zero-point motion of
$\mu^{+}$. In particular, one site is close to the FeAs layers
while the other one is much closer to the RE ions (see
Fig.~\ref{MuonLandscape}). These are referred to as ``A'' (or
``1'') and ``B'' (or ``2'') in the following, respectively, while
the unstable one is referred to as ``C''. For the sake of clarity,
the electrostatic landscape for LaFeAsO is reported in
Fig.~\ref{MuonLandscape} together with the crystallographic
positions of both stable and unstable sites. As a first-order
approximation, both $\zeta_{1}$ and $\zeta_{2}$ are
$T$-independent quantities. This is quite reasonable since the
diffusion of $\mu^{+}$ is typically a marginal process in the
explored $T$-range ($T \lesssim 170$ K). In the case of REFeAsO
materials, one typically finds $\zeta_{1}/\zeta_{2} \simeq 4$ for
the occupation probabilities of the sites $\zeta_{i}$,
independently from the actual chemical stoichiometry. The
lineshapes associated with the two different $\mu^{+}$ sites in
REFeAsO$_{1-x}$F$_{x}$ (RE = Ce) are narrow enough to resolve
signals from both of them in the undoped and in the slightly doped
($x = 0.03$) materials only \cite{Shi11}. A further increase in
$x$ yields a broadening of the frequency distribution, making the
smaller signal from site ``B'' unobservable. The interpretation of
the two contributions as signals coming from two inequivalent
sites is strongly supported by the experimental findings for all
the investigated samples. In particular, both signals show a fast growth either of the oscillation frequency
or of the damping at the same critical temperature
$T_{N}$, hence reflecting the same microscopic electronic
environment.

\begin{figure}[t!]
\vspace{8.2cm} \includegraphics{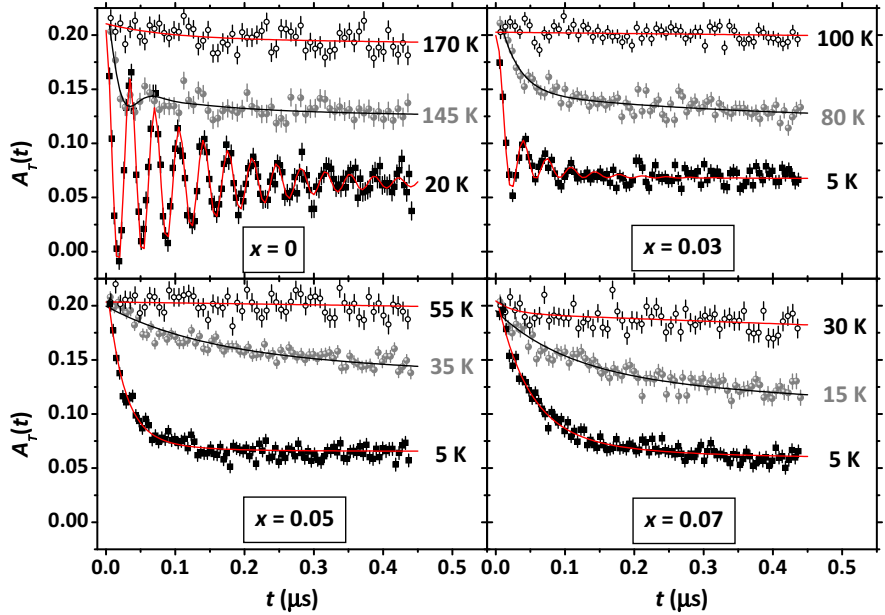} \caption{\label{DepFeCo}
Experimental spin-depolarization at low-$T$ in ZF conditions for
CeFe$_{1-x}$Co$_{x}$AsO in the low-$x$ region of the electronic
phase diagram  \cite{Pra13c}. The strong increase
of the transversal damping with increasing $x$ is the experimental
trademark of the progressive disordering of the magnetic phase
towards the limit of short-range magnetic order coexisting with
superconductivity.}
\end{figure}
\begin{figure}[t!]
\vspace{7.7cm} \includegraphics{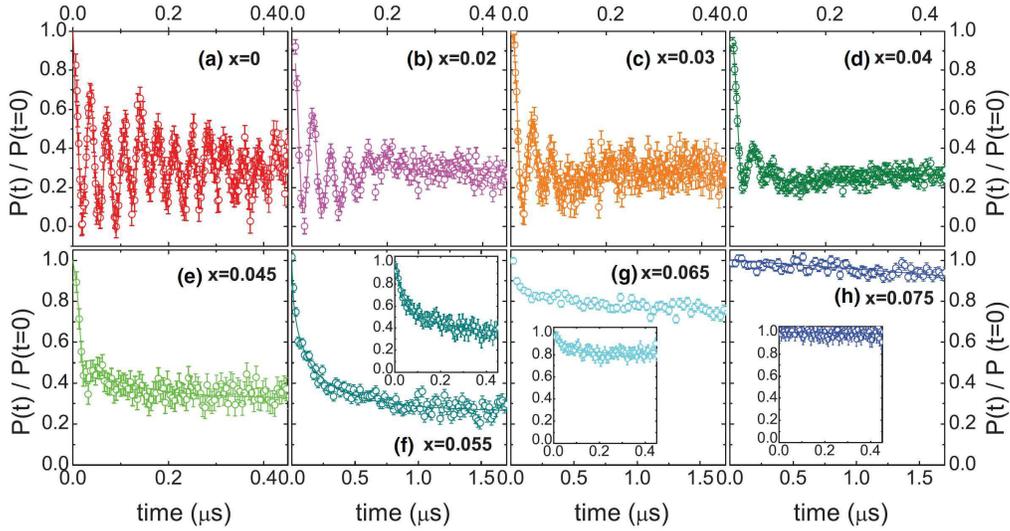} \caption{\label{DepBernhard} Experimental
spin-depolarization at low-$T$ in ZF conditions for
Ba(Fe$_{1-x}$Co$_{x}$)$_{2}$As$_{2}$ in the low-$x$ regime
(Figure reprinted from Ref. \cite{Ber12}. Copyright (2012) by the American Physical Society).}
\end{figure}

The $\mu^{+}$ implanted in both sites give rise to a
characteristic beat in the $\mu^+$ depolarization function of the
LaFeAsO parental magnetic compound at $T\ll T_N$ \cite{Kla08}. On the other
hand, this is not the case in compounds based on magnetic
rare-earth (RE) ions where only one oscillating signal is detected
from site ``1'' (see Fig.~\ref{DepFeCo}). In fact, site ``2''
being very close to the large RE magnetic moments which generate a
broad field distribution, shows an extremely fast signal decay
\cite{Shi11}. Both in the case of LaFeAsO and CeFeAsO, the
function $F_{1}(t) = \cos\left(\gamma_{\mu} B_{\mu_{1}} t\right)$
describes the data quite well ($\gamma_{\mu} = 2\pi \times 1.355
\times 10^{-2}$ MHz/G being the $\mu^{+}$gyromagnetic ratio). As
reported in Fig.~\ref{DepBernhard}, also in the case of the parent
compounds of the $122$ family (e.g. BaFe$_{2}$As$_{2}$) one
clearly observes beats arising from the signals of two muon sites
\cite{Ber12}.

When considering slightly charge-doped magnetic compounds,
long-range magnetism is still probed by $\mu^{+}$ even if some
qualitative changes in the spin-depolarization functions should be
considered. First of all, the oscillating cosine-like function
considered above in the case of the parent compounds should be
modified to $F_{1}(t) = J_{0}\left(\gamma_{\mu} B_{\mu_{1}}
t\right)$ ($J_{0}$ standing for a zeroth-order first-kind Bessel
function). This function better describes the field distribution
and the gradual modification of the spin density wave (SDW) phase
commensurability upon increasing the charge doping
\cite{Liu08,Car09}. On the other hand, the increase in the degree
of magnetic disorder yields to an enhancement of the transversal
damping, with respect to what observed in the parental compounds.
This is evident both in CeFeAsO$_{1-x}$F$_{x}$,
CeFe$_{1-x}$Co$_{x}$AsO (see Fig.~\ref{DepFeCo}) and in
Ba(Fe$_{1-x}$Co$_{x}$)$_{2}$As$_{2}$ (see Fig.~\ref{DepBernhard}).

By further increasing $x$, the disorder is so strong that the
oscillations are completely overdamped, hence reflecting an
extremely broad distribution of static magnetic fields at the
$\mu^{+}$ site. Even in this highly disordered configuration, one
can give an estimate of the typical internal field of the magnetic
phase by referring to the width of the field distribution (i.~e.,
the squared root of the second moment). It should be remarked that
in the case of overdamped transversal oscillations one should
carefully exclude dynamical effects on the muon depolarization
function. In these cases, one useful way to distinguish among the
two scenarios is to perform a longitudinal-field (LF) scan. Only
in the case of a static distribution of local fields the
application of a strong enough magnetic field allows one to quench
the spin-depolarization and to estimate the width of the static
distribution (see \cite{Shi11} for example).

In general, the three main microscopic contributions to the local
field at the muon site $B_{\mu}$ come from the dipolar, the
transferred hyperfine and the Lorentz terms (see \cite{Pra13b} and
references therein)
\begin{equation}\label{EqSumLocalFields}
B_{\mu} = \left|\textbf{B}_{dip}(\mathbf{r}_{\mu}) +
\textbf{B}_{c}(\mathbf{r}_{\mu}) + \textbf{B}_{L}\right|.
\end{equation}
The commensurate SDW actually corresponds to an  AFM configuration
and $B_{\mu_{1}}(T)$ is proportional to the sublattice
magnetization $M_{st}(T)$ (see \cite{DeR12}). Owing to the
crystallographic symmetry of site ``1'', for an AFM order the only
relevant contribution is the dipolar term, since the hyperfine
contribution is almost entirely averaged out and $\textbf{B}_{L}$
can be neglected because of the vanishing macroscopic
magnetization. Accordingly, if the crystallographic position of
site ``1'' is precisely known it is possible to estimate the Fe
magnetic moment from dipolar sums. From the internal field at the
$\mu^{+}$ site measured in the case of $1111$ materials with no
charge doping, one derives  $\mu_{Fe} \simeq 0.65 \; \mu_{B}$ for
the Fe magnetic moment \cite{DeR12}. Even in the case of a sizeable
degrees of magnetic disorder, the order parameter can be
estimated from the width of the distribution
of static local magnetic fields, as explained above.

\section{Electronic phase diagram and the coexistence between
magnetism and superconductivity}\label{SectionCoex}

The main typical outputs of ZF-$\mu^{+}$SR
experiments on magnetic materials are the order
parameter and the sample volume fraction $V_m$, which can be evaluated from Eq.~\ref{EqGeneralFittingZF} as two independent fitting
parameters. The great advantage of $\mu$SR when compared, for example, to neutron scattering is that it is a local probe
technique which can detect a static magnetic order even when the coherence length is reduced to a few lattice steps
and the system is quite inhomogeneous.
Thus, $\mu^{+}$SR is perfectly suited to investigate the details of the
transition  between the magnetic and the superconducting phases and it has been deeply employed
in pnictide superconductors to study the phase diagram obtained upon tuning in a controlled way some
key-parameter such as the level of iso- and/or hetero-valent chemical
substitutions, the structural parameters and the external pressure.

The crossover region between the magnetic and superconducting
electronic ground states is of crucial relevance for the
understanding of the intrinsic microscopic properties of pnictide
materials. According to some theoretical models
\cite{Fer10a,Fer10b}, the experimental finding of coexistence
between magnetism and superconductivity over a certain region of
the phase diagram is an indication of an unconventional pairing
among the supercarriers. Still, one should clarify which is the
spatial level of coexistence or, in other terms, the degree of
spatial intertwining of the two different order parameters. This
degree can be quantified by a characteristic length scale $d$
describing the typical distance among magnetic or, alternatively,
among superconducting domains. The strength of the claim of
``coexistence between magnetism and superconductivity'' is
crucially depending on the order of magnitude of $d$. In the
case $d \sim 100$ nm $- 1$ $\mu$m, one typically refers to the
so-called \emph{macroscopic or mesoscopic segregation} of the two
ground states. No definitive conclusion can be obtained in this
case concerning the symmetry of the superconducting order
parameter since the ground states occupy different spatial regions
that are in principle mutually independent from each other. This
condition can be driven for instance by trivial inhomogeneity of
the chemical doping in the investigated material and it has
typically to deal with \emph{extrinsic} properties. The opposite
case $d \sim 1 - 10$ nm is by far more interesting since it
implies a much finer intertwining of the two order parameters
(\emph{nanoscopic segregation} or \emph{nanoscopic coexistence})
towards the limit of the so-called ``\emph{atomic coexistence}''
ideally realized in the same spatial position. This is clearly a
much more interesting limit in order to draw conclusions on the
\emph{intrinsic} properties of the examined materials.

\begin{figure}[b!]
\vspace{7.8cm} \includegraphics{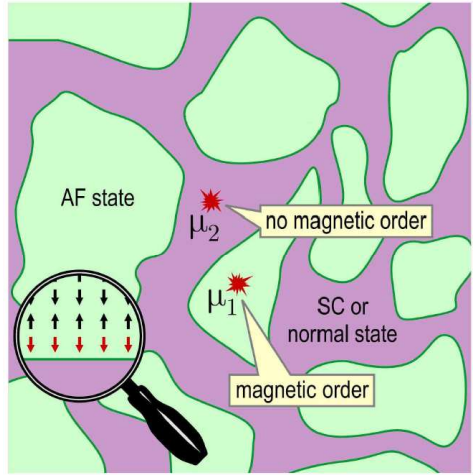} \caption{\label{SketchySegregation}(Color
online) Sketchy representation of the model of phase segregation of
magnetism and superconductivity (Figure reprinted from Ref. \cite{Par09}.
Copyright (2009) by the American Physical Society). The
nanoscopic coexistence is realized whether the order-of-magnitude of
the mean distance among magnetic domains (green droplets) is close
to $d \sim 1$ nm, namely the spatial range of the magnetic dipolar
field generated by uncompensated Fe moments on the domain walls (red
arrows).}
\end{figure}
The experimental way of treating this problem is typically not
trivial at all and qualitatively different results can be obtained
as a function both of the chemical homogeneity of the investigated
sample and of the technique employed. In this respect, it should
be stressed that $\mu^{+}$SR measurements are of crucial
importance for the reasons outlined above. The local nature of the
technique allows one to be sensitive to the disordered magnetism
that is typically realized in the case of nanoscopic coexistence.
As a further advantage with respect to other local techniques
like, e.~g. NMR, only $\mu^{+}$SR allows to precisely estimate
$V_{m}$. This information is of major importance in order to
quantify the order of magnitude of $d$. Let's refer to
Fig.~\ref{SketchySegregation} where a sketchy representation of
the phase segregation of two different order parameters is
reported \cite{Par09}. The global magnetic moment within the domains
is null due to the AFM correlations characteristic of the SDW
phase. Still, $\mu^{+}$ are able to feel the dipolar field
generated by the uncompensated magnetic moments on the domain
walls (see red arrows in the figure under the magnifying lens). In
the case of the dipolar field generated by Fe in the case of
$1111$ oxy-pnictide materials, one can roughly deduce that the
minimal distance required in order to probe magnetism under these
conditions for $\mu^{+}$ implanted out of the domains is of the
order of $1$ nm \cite{San10}. In the case of the mesoscopic
segregation of the order parameters, $\mu^{+}$ implanted out of
the magnetic domains (and conventionally labelled as
$\mu_{2}^{+}$) are on the average too far away from the domains
themselves to probe static magnetism and only $\mu^{+}$ implanted
\emph{into} the domains (and conventionally labelled as
$\mu_{1}^{+}$) give rise to a magnetic signal. As a result, $V_{m}
< 1$ in the case of mesoscopic segregation and a value $\left(1 -
V_{m}\right) > 0$ is measured as a paramagnetic volume fraction,
accordingly. This behaviour was clearly enlightened in the case of
La-based $1111$ samples with the phase diagram being swept both by
the increasing concentration of electrons triggered by the
O$_{1-x}$F$_{x}$ substitution \cite{Pra13a} or by the application
of external hydrostatic pressure \cite{Kha11}, at variance with
what is typically reported for $1111$ materials based on magnetic
RE ions, as it will be shown subsequently.
\begin{figure}[t!]
\vspace{10.7cm} \includegraphics{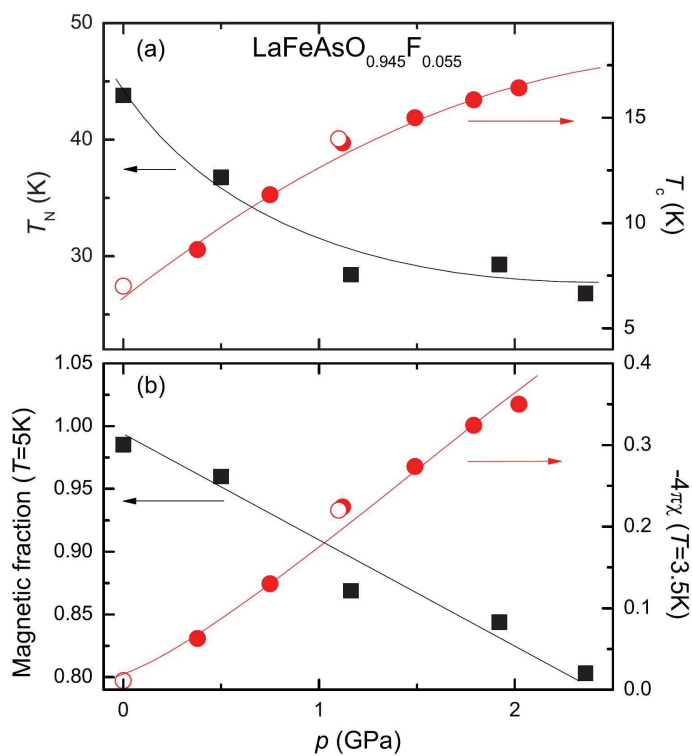}
\caption{\label{SegregationKhasanov} Effect of pressure ($P$) on
magnetism and superconductivity in LaFeAsO$_{0.945}$F$_{0.055}$
 \cite{Kha11}. Upper panel: $P$-dependence of the
magnetic and superconducting critical temperatures. Lower panel:
$P$-dependence of the magnetic and superconducting volume
fractions.}
\end{figure}
At the same time, early reports on hole-doped $122$ compounds like
Ba$_{1-x}$K$_{x}$Fe$_{2}$As$_{2}$ seemed to confirm the picture of
the mesoscopic segregation \cite{Par09} at variance with what
reported about electron-doped compounds like
Ba(Fe$_{1-x}$Co$_{x}$)$_{2}$As$_{2}$ \cite{Ber12,Lap09,Pra09}. It
should be remarked that independent quantifications of the
superconducting shielding fractions of samples displaying
mesoscopic segregation (e.g. derived from dc magnetometry or ac
susceptibility) show a complementary trend with respect to that of
the magnetic volume fraction across the phase diagram. In
particular, a slight decrease in the saturation value of $V_{m}$
induced by the increase of $x$ in
La$_{0.7}$Y$_{0.3}$FeAsO$_{1-x}$F$_{x}$ \cite{Pra13a} or by the
increase of external hydrostatic pressure in
LaFeAsO$_{1-x}$F$_{x}$ \cite{Kha11} is typically reflected into a
specular increase in the saturation value for the superconducting
shielding fraction. This is clearly explained by
Fig.~\ref{SegregationKhasanov} where these quantities are reported
for LaFeAsO$_{0.945}$F$_{0.055}$ as a function of the external
pressure. These results confirm the picture of segregation of the
two different electronic environments into well-separated regions
strongly competing for volume. It should be remarked that such
picture for LaFeAsO$_{1-x}$F$_{x}$ is fully consistent with the
first-order-like transition between the two electronic ground
states reported in \cite{Lue09}.

\begin{figure}[t!]
\vspace{8.2cm} \includegraphics{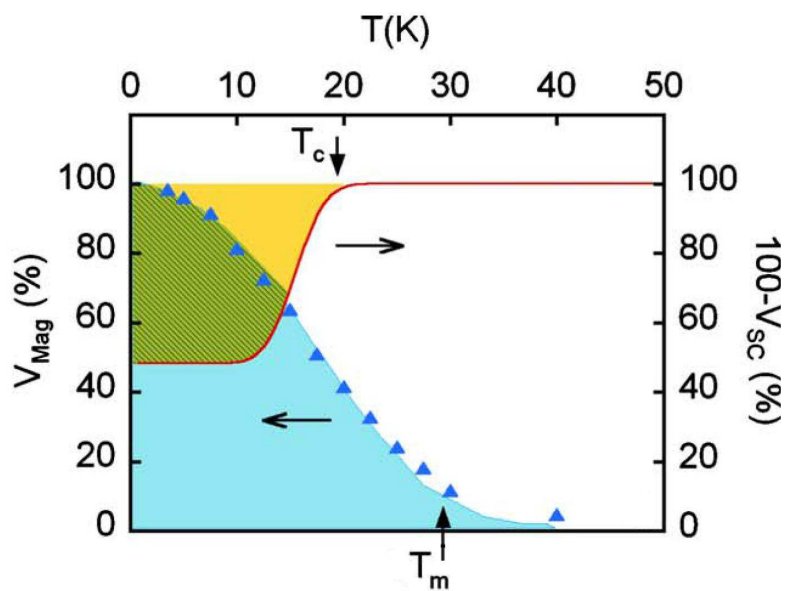} \caption{\label{CoexSam} Nanoscopic
coexistence of magnetism and superconductivity in
CeFeAsO$_{0.94}$F$_{0.06}$  \cite{San10}.}
\end{figure}
On the other hand, the cases of Ce- and Sm-based $1111$
oxy-pnictides were shown to be dramatically different from what
observed in LaFeAsO$_{1-x}$F$_{x}$. In these materials, the
coexistence region of magnetism and superconductivity was shown to
be characterized by $V_{m} \simeq 100$ \% with a bulk fraction of
the sample being at the same time superconductor, as illustrated
in Fig.~\ref{CoexSam} in the case of CeFeAsO$_{0.94}$F$_{0.06}$
\cite{San10}. These findings are interpreted by coming back to
Fig.~\ref{SketchySegregation} and by now assuming that $d \sim 1
- 10$ nm. Under these circumstances, both $\mu_{1}^{+}$ and
$\mu_{2}^{+}$ (namely, all the implanted $\mu^{+}$) probe static
magnetism being the maximum distance between different magnetic
domains of the same order of magnitude of $d$, namely the spatial
range for the dipolar fields generated by domain walls.
Accordingly, one has $V_{m} \simeq 100$ \% and it must be assumed
that the interstitial regions between the different magnetic
domains are superconducting since dc magnetometry confirms that a
bulk fraction of the sample contributes to the diamagnetic
shielding. Such fine intertwining of different order parameters
was detected in SmFeAsO$_{1-x}$F$_{x}$ \cite{San09a,Dre08,Dre09},
in CeFeAsO$_{1-x}$F$_{x}$ \cite{San10,Shi11}, in
CeFe$_{1-x}$Co$_{x}$AsO \cite{Pra13c}.
\begin{figure}[t!]
\vspace{10.7cm} \includegraphics{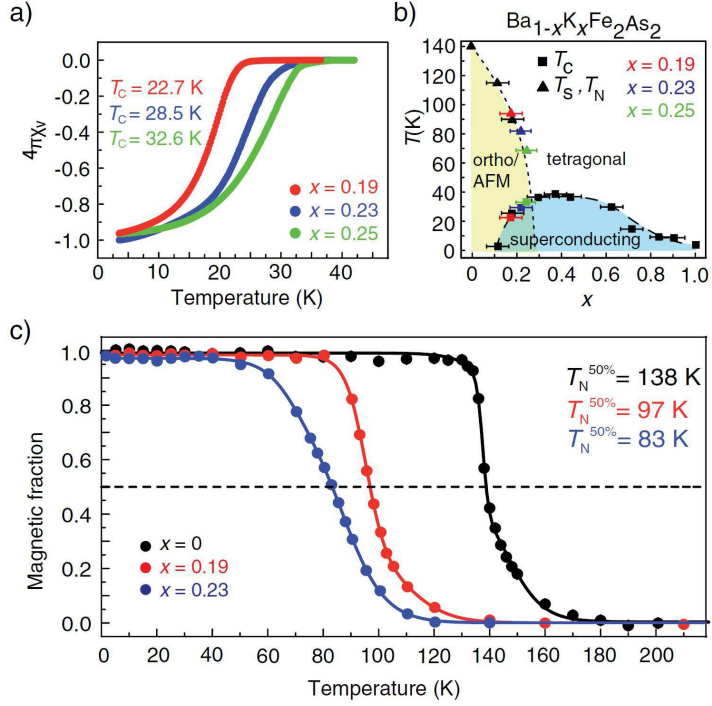} \caption{\label{CoexLue} Nanoscopic
coexistence of magnetism and superconductivity in
Ba$_{1-x}$K$_{x}$Fe$_{2}$As$_{2}$ (Figure reprinted from Ref. \cite{Wie11}.
Copyright (2011) by the American Physical Society).
Panel (a): superconducting shielding fractions. Panel (b):
electronic phase diagram for Ba$_{1-x}$K$_{x}$Fe$_{2}$As$_{2}$.
Panel (c): magnetic volume fractions.}
\end{figure}

\begin{figure}[t!]
\vspace{9.7cm} \includegraphics{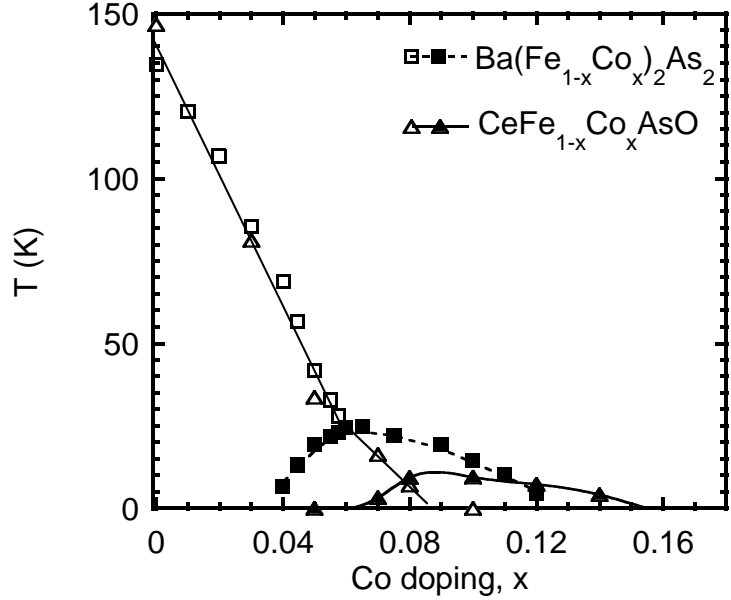} \caption{\label{PhCo}
The phase diagram of Ba(Fe$_{1-x}$Co$_x$)$_2$As$_{2}$ \cite{Ber12} and of CeFe$_{1-x}$Co$_x$AsO \cite{Pra13c} are reported as a function of the Co content $x$.
Open symbols show the behaviour of $T_N(x)$ while the closed symbols show that of $T_c(x)$.
}
\end{figure}

It should be remarked that
the picture of mesoscopic segregation described above for the case
of Ba$_{1-x}$K$_{x}$Fe$_{2}$As$_{2}$ was modified after the
investigation of samples with higher chemical homogeneity,
displaying that the nanoscopic coexistence is a more suited
framework for such materials \cite{Wie11}. At the sake of clarity,
results relative to the hole-doped
Ba$_{1-x}$K$_{x}$Fe$_{2}$As$_{2}$ are reported in
Fig.~\ref{CoexLue}  \cite{Wie11}. It is rather interesting to
observe that a static magnetic ground-state is recovered also by
diluting optimally electron-doped $1111$ materials with Ru
\cite{San11,San13}. Remarkably, the magnetic and superconducting
ground-states coexist at the nanoscopic level and the latter one
is completely suppressed for a Ru content around 60\% (see Fig.
5). The observation of such a tiny effect of diamagnetic
impurities on the superconducting transition temperature would
appear at first in contrast with a magnetic pairing mechanism and
has lead to alternative explanations, as the ones involving
orbital currents \cite{Kon10,Ona12}. Still, it is possible that
the subtle interplay between intraband and interband pairing
processes can give rise to such a slow decrease of $T_c$ even in
the framework of a magnetic pairing mechanism \cite{Vav11,Fer12}.
Finally, it is important to point out that the different families of IBS show some significant differences.
In Fig.~\ref{PhCo} the phase diagram of  Ba(Fe$_{1-x}$Co$_x$)$_2$As$_{2}$ \cite{Ber12} and of CeFe$_{1-x}$Co$_x$AsO \cite{Pra13c} is reported.
In spite of the quantitatively similar trend of the magnetic ordering temperature, which in the Ce-based 1111 compound was shown to be long-range
for $x<0.03$ and short-range at higher doping levels \cite{Pra13c}, one notices clear differences at the crossover between the magnetic
and the superconducting ground-states.  In the 122 compound the coexistence region between the two phases is quite extended
while in the 1111 compound it is only marginal. Moreover, the superconducting transition temperature is lower in the 1111 compounds.
This suggests that the differences in the band structure, namely in the FS nesting, and in the anisotropy of the two family of compounds could
play a relevant role in determining the superconducting properties, particularly when the charge doping is associated with the introduction
of impurities in the FeAs planes.

\section{Transverse-field $\mu^+$SR in superconducting
materials}\label{SectionTF}

\begin{figure}[t!]
\vspace{10.7cm} \includegraphics{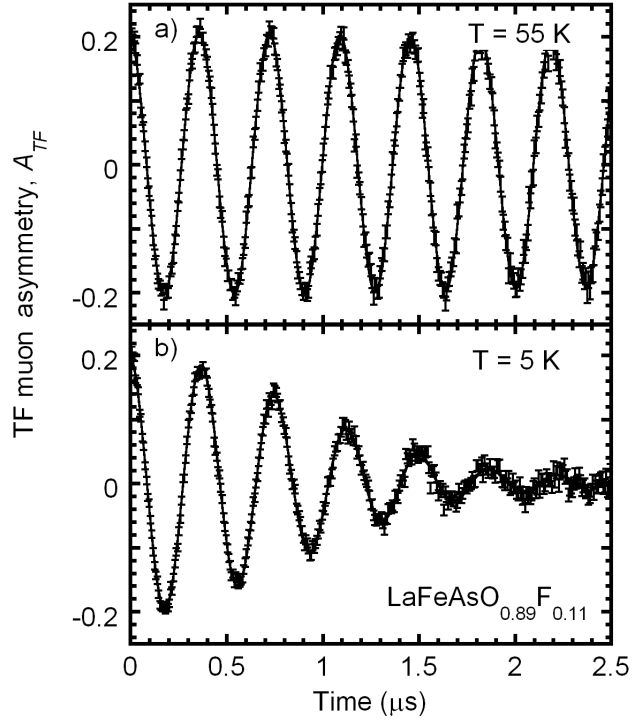} \caption{\label{TFspectra}
Representative time evolution of the TF-$\mu$SR asymmetry for a
polycrystalline sample of LaFeAsO$_{0.89}$F$_{0.11}$ at $\emph{T}$
values well above and below the superconducting transition
temperature $T_c=28$ K, performed in an external magnetic field
$H_0=200$ G perpendicular to the initial muon spin
polarization. Solid curves are best fits to a Gaussian damped
precession (see text).}
\end{figure}

When type II superconductors, as the IBS, are field-cooled in an
external field $H_{c1}< H_0< H_{c2}$ ($H_0, H_{c1}$ and $<H_{c2}$
being the external, the lower and the upper critical field,
respectively) the magnetic field is not homogeneously distributed
throughout the sample but concentrated along tubes which form a
FLL. Since the distance among two adjacent flux lines is much
larger than the distance between the muon sites, the muons are
able to finely probe the field distribution generated by the FLL.
Such distribution leads to an enhancement of the muon
depolarization rate which can be suitably probed in transverse
field (TF) $\mu^+$SR experiments. The FLL field distribution is
not symmetric around $H_0$ \cite{Bra95} since it reflects the
minimum, the maximum and the saddle-point values of the magnetic field
profile within the triangular FLL unit cell. In real samples, and
in particular in powders, the FLL imperfections arising from
randomly distributed pinning centers average out these asymmetries
and the field distribution tends to a Gaussian one, centered
around the average internal field $B_{\mu}$. Hence, the TF muon depolarization function can be
approximated by $D_{TF}(t) \approx
exp[-(\sigma_{SC}^2 + (\sigma^s_N)^2)t^2/2]$, with $\sigma^s_N$ the weak
relaxation due to the nuclear magnetic dipoles (see Eq. 2) and
$\sigma_{SC}$ the second moment of the FLL field distribution,
namely $\sigma_{SC}/\gamma_{\mu}=\!\!=\!\!(\overline{B}^2-
{B_{\mu}^2})^{1/2}$. The latter term is proportional to the
inverse square of the London penetration depth $\lambda_L$ and, in
turn, to the supercarrier density $n_s$. Accordingly, one can
write that $\sigma_{SC}(T)\propto n_s/m^*$ with $m^*$ the electron
effective mass. Since each flux line is surrounded by a screening
current over a distance of the order of $\lambda_L$, the average
field at the muon site $B_{\mu}$ is slightly lower than the
applied field, namely $B_{\mu}= (1+4\pi\chi)H_0$ with $\chi<0$.

\begin{figure}[t!]
\vspace{11.2cm} \includegraphics{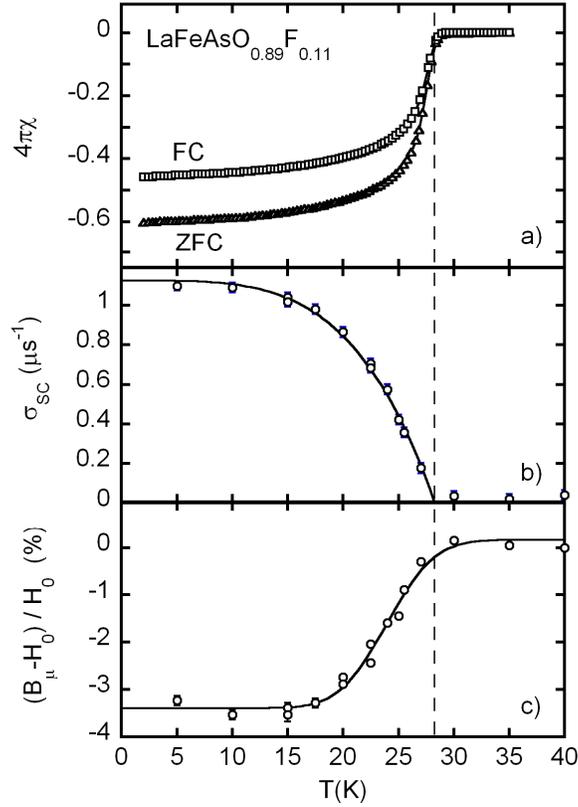}
\caption{\label{TFparameters} Superconducting response of
LaFeAsO$_{0.89}$F$_{0.11}$ as a function of temperature. \emph{a})
Zero field cooled (ZFC) and field cooled (FC) magnetic
susceptibility, evaluated from magnetization measurements
performed at $H= 5$ Gauss by assuming a nearly spherical grain
powders with demagnetization factor $N=4\pi/3$. \emph{b}) Relaxation
rate $\sigma_{SC}$ fitted using a standard BCS curve; the constant
behavior at low T is indicative of a gap function without nodes.
\emph{c}) Relative shift of the magnetic field at the muon site
$B_\mu$ respect to the applied field $H_0$. The lines in panel a)
and c) are guides for the eye.}
\end{figure}

Fig.~\ref{TFspectra} shows the typical time evolution of the TF
muon asymmetry $A_{TF}(t)$ in LaFeAsO$_{0.89}$F$_{0.11}$, measured
at $T$ values above (a) and below (b) $T_c$= 28 K. The
superconducting critical temperature was derived from the
$T$-dependence of the magnetic susceptibility, $\chi(T)$, as shown
in Fig~\ref{TFparameters}(a). The $A_{TF}(t)$ data fit rather well
to the function
\begin{equation}\label{EqTF1}
\frac{A_{TF}(t)}{A_{TF}(0)}= a_{TF}\cdot exp[-(\sigma_{SC}^2+\sigma_N^2)t^2/2] cos
(\gamma_{\mu} B_\mu t)
\end{equation}
throughout the whole temperature range $0< T< 50$ K with a constant
amplitude $a_{TF}$. This ensures that all the muons probe the same
field distribution, i.e. the sample
is always single phase and fully superconducting. %The discrepancy with the value at zero temperature, $\chi_0$,
The depolarization rate for $T>T_c$ is Gaussian and its rather low
value, $\sigma_N= 0.17 \mu s^{-1}$, indicates that the amount of
dilute magnetic impurities, which are often present in IBS %mainly as Fe$_3$O$_4$
and cause an unwanted enhancement in the relaxation rate
\cite{Biswas2010,San09a}, is negligible in that sample. The
temperature evolution of the FLL contribution to the
depolarization rate, $\sigma_{SC}(T)$, and the relative shift of
the field at the muon site, $\left( B_\mu(T)-H_0\right)/H_0$, due
to the diamagnetic shielding are shown in
Fig.~\ref{TFparameters}(b) and (c), respectively. The development
of the FLL is clearly evidenced both by the increase of
$\sigma_{SC}$ and by the diamagnetic shift of $B_\mu$ below
$T_c^{TF}=28 K$, which coincide with that measured by $\chi(T)$.
From the value of $\sigma_{SC}$ extrapolated at T=0, by using the
relation $\sigma_{SC}=7.58\times 10^{-4}\lambda_L^{-2}$, in CGS
units, \cite{Barf1988,San2008}, the London penetration depth is
estimated to be $\lambda_L\simeq 2620$ \AA. The temperature dependence
of $\sigma_{SC}(T)$ follows the s-wave weak coupling BCS
temperature dependence predicted by the two fluid model
$\sigma_{SC}(T)\propto 1/\lambda^2(T)\propto [1-(T/T_c)^4]$, as
shown by the curve in Fig.~\ref{TFparameters}(b). In particular,
the flat behavior for $T< T_c/3$ is indicative of the absence of
nodes in the gap function.

\begin{figure}[t!]
\vspace{8.7cm} \includegraphics{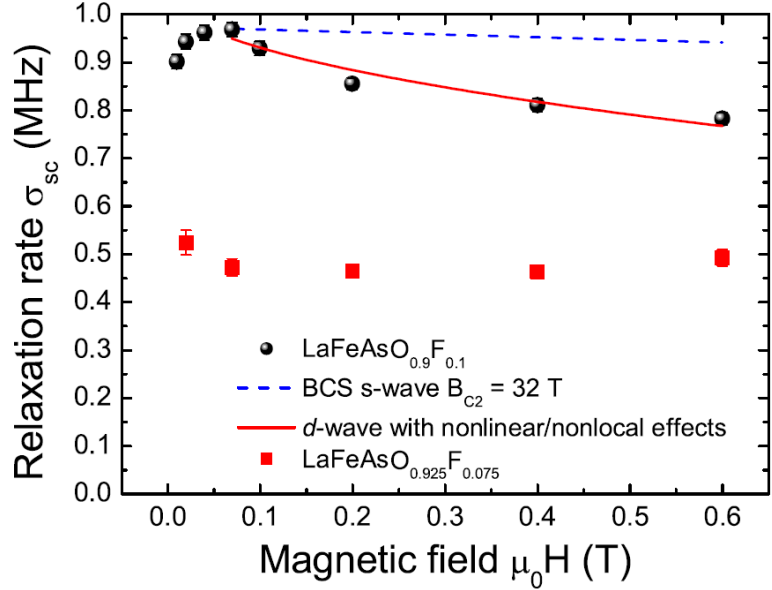} \caption{\label{sigmavsH} Field
dependence of the muon relaxation rate $\sigma_{SC}$ at $T= 1.6$ K
for LaFeAsO$_{0.925}$F$_{0.075}$ and LaFeAsO$_{0.9}$F$_{0.1}$
powder samples with data fit to a d-wave (solid line) or s-wave
(dashed line) behavior (Figure reprinted from Ref. \cite{Luetkens2008}. Copyright (2008)
by the American Physical Society).}
\end{figure}

\begin{figure}[t!]
\vspace{9.5cm} \includegraphics{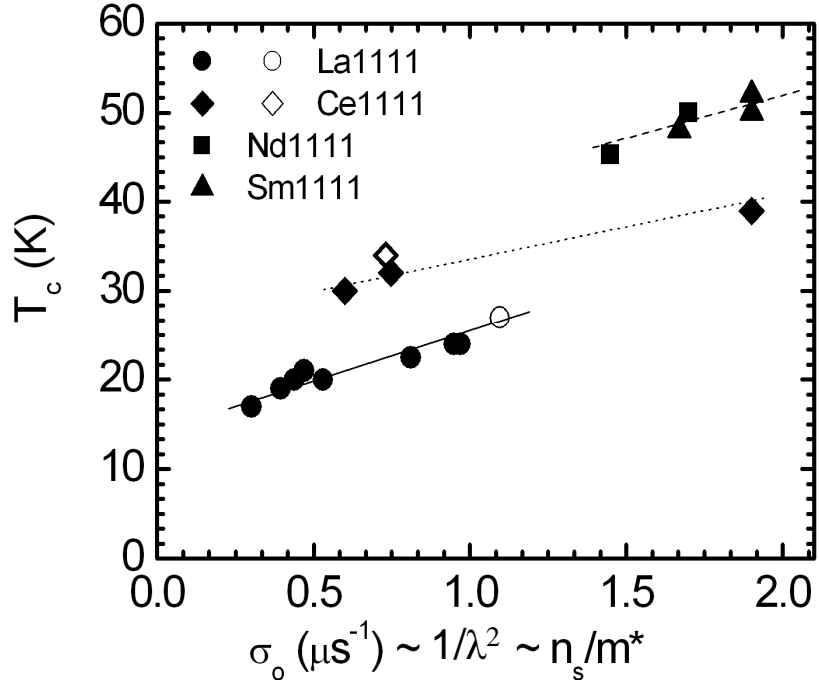} \caption{\label{uemura} Uemura
plot for 1111 families with different rare earth ions, with $\sigma_{SC}$ values
measured in a transverse field $H_0\sim B^{max}$ (see text). Open
and solid symbols are from this work and from Refs.
\cite{Luetkens2008,Lue09,Carlo2009,Khasanov2008,Takeshita2009,
Dre08, Maeter2013}, respectively.}
\end{figure}

The s-wave-like $T$-dependence of $\sigma_{SC}$ was also observed
in a La-based 1111 compound with $T_c=23$ K \cite{Luetkens2008}.
However, this behavior is in conflict with the observed dependence
of $\sigma_{SC}$ on the magnitude of the applied field $H_0$. In
fact, as it is shown in Fig.~\ref{sigmavsH} \cite{Luetkens2008},
$\sigma_{SC}$ displays a maximum for a field value $B^{max}\approx
200\div2000$ G and at higher fields the relaxation rate decreases,
which is usually indicative of superconductivity with nodes in the
gap function \cite{Luetkens2008, Sonier2000}. It should be noticed
that this feature, also found in Sm-based 1111 compounds, could be
rather due to the presence of different superconducting gaps
\cite{Weyeneth2010}, as it can be expected for system with several
bands crossing the Fermi level. This field dependence should be
carefully considered when comparing results from different
experiments in the so-called Uemura like plot, where $T_c$ is
plotted as a function of $\sigma_{SC}(T=0) \propto n_s(T=0)$. In
Fig.~\ref{uemura} this behavior is shown for the 1111 family as a
function of the F doping for those values of $\sigma_{SC}(T=0)$
measured at a field close to $B^{max}$. Although a theoretical
description of the Uemura plot for the 1111 IBS is still missing,
it is clear that the rate of the decrease of $T_c$ with $n_s(0)$
strongly depends on the symmetry of the order parameter, on the
presence of nodes in the gap and on the occurrence of pair
breaking processes. Hence, a suitable modeling of that plot would
have significant implications in the clarification of the pairing
mechanism for IBS. Nevertheless, it has also to be remarked that a
temperature dependence of $\sigma_{SC}$ can be observed at low $T$
in underdoped La-based 1111 samples \cite{Takeshita2009}, leading
to some uncertainty in the estimate of $\sigma_{SC}(T=0)$. Indeed,
while analyzing the behavior of $\sigma_{SC}(T)$ a great care must
be taken in order to discern those effects not associated with the
$n_s$ $T$-dependence, as it will be described in the following
section.

\section{Transverse-field $\mu^+$SR in case of coexistence between
magnetism and superconductivity}\label{SectionTFcoex}

In IBS, two main effects might alter the TF depolarization rates.
One is extrinsic and comes from diluted ferromagnetic impurities
which are often present in polycrystalline samples. These can be
recognized in ZF experiments by an exponential depolarization
either of the TF-muon asymmetry for $T>T_c$ or of the ZF-muon
asymmetry spectrum for $T>T_N$. In TF-$\mu^+$SR experiments, this
exponential term sometimes can be subtracted after a calibration
above $T_c$ and kept constant over the whole temperature range,
even if this procedure might be affected by some error due to the
field/temperature dependence of the spurious magnetic
contribution. Another important effect might originate from the
presence of a static order or of magnetic correlations developing
close to it, especially in underdoped compounds or when magnetic
rare earths are present. In particular, when a static order
appears the corresponding spontaneous internal fields $B_i$ add to
$H_0$ and the muons detect a local field
$B_{mag}\approx|$\textbf{H$_0$}+ \textbf{B$_i$}$|$. Hence both the
value of the internal fields and that of the depolarization rate
are strongly enhanced. Their behavior depends on the distribution
width of the internal fields $B_i$ which, in principle, requires a
vectorial analysis to take into account the muon spin projection
along the detector direction \cite{Allodi2006}. Under those
circumstances, the information on the superconducting state is
usually lost.

\begin{figure}[t!]
\vspace{11.7cm} \includegraphics{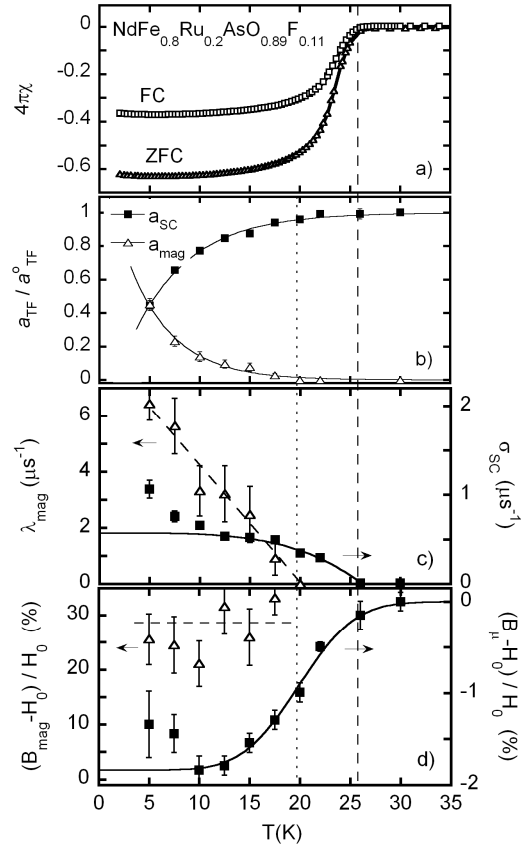}
\caption{\label{TFparacoex} Zero field cooled (ZFC) and field
cooled (FC) magnetic susceptibility, evaluated from magnetization
measurements performed at $H= 5$ G by assuming a nearly
spherical grain powders with demagnetization factor $N=1/3$.
\emph{b}) Muon fractions $a_{mag}(T)$, probing nanoscopic regions
with sizeable internal fields $B_i$ from the spontaneous Fe
magnetic order, and $a_{SC}(T)$, where the superconducting FLL
distribution is almost unperturbed. \emph{c}) Relaxation rate of
the muon asymmetry and \emph{d}) relative shift of the magnetic
field at the muon site respect to the applied field $H_0$, for the
magnetic (left axis) and non magnetic (right axis) contributions.
All the lines are guides for the eye.}
\end{figure}

In the 1111 compounds, where magnetism and superconductivity
coexist at the nanoscopic level, $B_i$ ranges from a maximum field
of few hundreds of Gauss to nearly zero, since it originates from ordered
moments varying over distance from less than one to four-five
lattice steps. In this case, one may assume that a fraction of the
muons experiences internal fields lowered by $4\pi|\chi|H_0$.
Then, the typical TF signal of the FLL is roughly distinguishable,
even if it is still slightly affected by the nearby $B_i$ which
give rise to a moderate increase  both of $B_\mu(T)$ and of
$\sigma_{SC}(T)$. For example, here the behavior of a
NdFe$_{0.8}$Ru$_{0.2}$AsO$_{0.89}$F$_{0.11}$ powder sample is
considered. This sample has $T_c=26$ K (Fig~\ref{TFparacoex}(a))
and by ZF muon experiments it was shown to display a magnetic
transition with an onset at $T_N^{onset}\approx20$ K and a full
magnetic volume fraction $V_m\simeq 1$ below $T= 5$ K, with
internal fields at the muon site of the order of 200 G (see Fig. 1
of Ref. \cite{San13}). The TF asymmetry data (not shown) fit quite
well to the sum of two oscillating terms
\begin{eqnarray}\label{EqTF2}
\frac{A_{TF}(t)}{A_{TF}(0)} & = & a_{SC} \cdot exp(-\Lambda t)\cdot
exp[-(\sigma_{SC}^2)t^2/2]\cdot cos (\gamma_{\mu} B_\mu t)\nonumber\\
& + & a_{mag}\cdot exp[-\lambda_{mag} t]\cdot cos(\gamma_{\mu} B_{mag}
t)\,\,\, ,
\end{eqnarray}
where the first accounts for those regions with vanishing $B_i$,
where the field distribution still reflects the FLL for
$T>T_N^{onset}$, while the second one accounts for those regions
where the spontaneous internal field is of the order of $H_0$. The
$\Lambda$ decay rate takes into account the exponential decay
measured already at high temperature, close to the value
$\Lambda\simeq 0.8 \mu s^{-1}$ and nearly temperature independent, which
is most probably due to the presence of a small fraction of
diluted magnetic impurities. The temperature dependence of the TF
parameters is shown in Fig.~\ref{TFparacoex}. Superconductivity is
well reflected by the behavior both of the increase of
$\sigma_{SC}(T)$ and of the diamagnetic shift
(right axis of panels c and d, respectively) below the same $T_c$
determined by the $\chi(T)$ curve. At low temperature they both
deviate from the behavior expected for a single SC phase compound,
roughly for $T<10$ K where the ZF experiment displays a sizeable
magnetic volume fraction \cite{San13}, confirming the nanoscopic
character of the coexistence. The panel Fig~\ref{TFparacoex}(b)
shows that the $a_{SC}/a^0_{TF}$ fraction is close the unity in
the temperature range $T_N^{onset}<T<T_c$, namely the full
sample is in the superconducting phase. Below $T_N^{onset}\approx
20$ K, which agrees with the ZF measurements reported in Fig. 1 of Ref.
\cite{San13}, the regions where the spontaneous fields sum to the
external one are signaled by the increase of the muon fraction
$a_{mag}$ displaying the field at the muon site $B_{mag}>H_0$
(left axis of Fig~\ref{TFparacoex}(d)) and high values of the
depolarization rate $\lambda_{mag}$ (left axis of
Fig~\ref{TFparacoex}(c)) which grows according to the
increase of the magnetic volume fraction.

\section{Summarizing Remarks}

Only five years after the discovery of superconductivity in the Fe
pnictides, $\mu^+$SR experiments have allowed to reach several
relevant milestones in the understanding of the microscopic
properties of IBS. As shown in Sect. 3, thanks to its sensitivity
to the magnetic order, even when it is short range, $\mu^+$SR has
played a crucial role in the description of the phase diagram of
the IBS, in the understanding of the phase transition between the
magnetic and the superconducting phases and of their possible
coexistence at the nanoscopic level. The TF-$\mu^+$SR studies
reported in Sect. 4, performed in the mixed phase where the FLL
sets in, have provided a description of the evolution of the
superconducting carrier density $n_s$ as one spans through the
phase diagram of IBS and shown how this quantity is correlated
with $T_c$. The behaviour of $n_s(T)$ and of $n_s(T\rightarrow 0)$
vs $T_c$ have given relevant hints on the symmetry of the
superconducting order parameter and on the microscopic mechanisms
underlying the superconductivity in IBS. A remarkable experimental
effort is still required to have a clear understanding of the
physics of IBS and to distinguish the peculiarities of each family
of compounds from the relevant features underlying a common
description of the superconducting condensate.

\section{Acknowledgements}

We would like to thank all those colleagues who have recently
collaborated with us in the realization of the $\mu$SR experiments
in the IBS. In particular: Alex Amato, Pietro Bonfà, Lucia Bossoni, Sean Giblin, Rustem
Khasanov, Gianrico Lamura, Hubertus Luetkens, Marcello Mazzani and
Toni Shiroka. Pietro Carretta and Samuele Sanna also acknowledge
the financial support from Fondazione Cariplo (research grant no.
2011-0266) for  the research activity on IBS. Giacomo Prando
acknowledges support from the Leibniz-Deutscher Akademischer
Austauschdienst (DAAD)Post-Doc Fellowship Program

%%%%%%%%%%%%%%%%%%%%%%%%%%%%%

%%%%%%%%%%%%%%%%%%%%%%%%%%%%%%%%%%%%

\end{document}